\documentclass[twocolumn]{aastex631}
\begin{document}

% for float placement:
\renewcommand{\topfraction}{1.0}
\renewcommand{\bottomfraction}{1.0}
\renewcommand{\textfraction}{0.0}

\shorttitle{Performance of the Southern Astrophysical Research Telescope Speckle Instrument}
%\shortauthors{Tokovinin et al.}

\title{Performance of the SOAR Speckle Instrument}

\author{Andrei Tokovinin}
\affil{Cerro Tololo Inter-American Observatory --- NFSs NOIRLab Casilla 603, La 
Serena, Chile}
\email{andrei.tokovinin@noirlab.edu}

\begin{abstract}
The  High  Resolution Camera  (HRCam)  speckle  imager  at the  4.1  m
Southern  Astrophysical  Research  telescope is  a  highly  productive
instrument that has  accumulated about 40K observations  to date.  Its
performance   (detected   flux,   level   of   the   speckle   signal,
signal-to-noise ratio, and limiting magnitude)  is studied here using both
the  actual data  and  realistic simulations,  including the  detector
noise.   In the  calculation  of the  speckle  power spectrum,  signal
clipping  is essential  to reduce  the noise  impact and  maximize the
sensitivity.   Increasing exposure  time of  individual frames  beyond
30\,ms does not improve the limiting magnitude, which ranges from 11.5
to 14  mag under a seeing  from 1\farcs6 to 0\farcs6  in the wide-band
$I$  filter. A  gain of  at  least one  magnitude is  expected if  the
current electron  multiplication CCD  is replaced  by a  high-end CMOS
detector with  a subelectron readout  noise.  This study will  help in
planning, executing,  and automating future speckle  observations with
HRCam and other speckle imagers.
\end{abstract} 

%---------------------------------------------------------
\section{Introduction}
\label{sec:intro}

Speckle interferometry, introduced by \citet{Labeyrie1970}, has become
a  standard   method  of   high-resolution  observations   at  optical
wavelengths, complementing  adaptive optics in the  infrared.  At that
time, recording and processing of a  large number of images with short
exposure  and  fine  pixel  scale  presented  a  formidable  technical
challenge.   With   the  advent  of  modern   computers  and  electron
multiplication (EM) CCDs, however, speckle interferometry has become a
technically        feasible       and        efficient       technique
\citep{Cantarutti2008,Horch2009}.   Nowadays, appearance  of low-noise
CMOS  detectors challenges  the dominant  role of  EM CCDs  in speckle
interferometry and opens new horizons.  The principles of speckle
  interferometry are covered in numerous textbooks and papers, e.g. by
  \citet{Goodman1985}.

Speckle interferometry serves mostly for  the study of orbital motions
of  binary and  multiple stars,  for  surveys of  binarity in  various
stellar  samples,   and  for  screening  exoplanet   hosts  for  close
companions.   Speckle instruments based  on EM CCDs are  in active
  use, e.g.  the DSSI \citep{Horch2009,Davidson2024}, the twin speckle
  cameras at Gemini \citep{Scott2018,Scott2021}, the  Quad speckle camera at the
  4 m Lowell Discovery Telescope  \citep{Clark2024a}, and the speckle
  instrument of the 6 m telescope \citep{Mitrofanova2021}.  The Gaia
mission \citep{Gaia1,EDR3} is a powerful  driver of current and future
speckle programs that provide an essential complement to Gaia.

The  High Resolution  Camera (HRCam)  has  been used  at the  Southern
Astrophysical Research (SOAR) 4.1 m telescope in Chile since 2007.  It
has been a very productive  and low-maintenance instrument for speckle
interferometry  \cite[see][and references  therein]{SAM22}.  In  2017,
the  detector was  upgraded to  an  iXon EM  CCD with  a high  quantum
efficiency and low noise \citep{HRCAM}.   HRCam uses 10--15 nights per
year and typically covers 300 targets in a night.  To date (2024 May),
there are  40,776 accumulated  speckle observations of  16,319 targets
(measurements  of binary  stars  and nonresolutions)  ---  one of  the
largest  speckle  data sets  in  the  world.  A  brief review  of  the
scientific use of the HRCam data can be found in \citet{HRCAM}.

So  far,  the performance  of  HRCam  received little  attention;  all
efforts have  been devoted to the  observations.  Our goal here  is to
quantify the quality of the speckle data by such parameters as speckle
contrast and signal-to-noise ratio (SNR) and to study their dependence
on the  observing conditions and instrument  parameters.  Estimates of
the expected SNR will help  in planning future observations by setting
acceptable conditions  for each  target. Eventually, the  data quality
prediction  and control  will  enable automation  of the  observations,
which, so far, largely depend on the observer's expertise.  An additional
motivation is to evaluate the potential  gain in sensitivity offered by an
upgrade to a high-end CMOS detector.

In  Section~\ref{sec:perf},  the  requisite information  on  HRCam  is
assembled, including the detector parameters, data processing, and 
photometric     calibration.     Simulations     are    covered     in
Section~\ref{sec:simul},  quantifying the  impact  of finite  exposure
time and spectral bandwidth and the role of the detector noise.  The simulation
code  is described  in the  Appendix.  Based on  the simulations,  the
benefit   of   upgrading  to   a   CMOS   detector  is   explored   in
Section~\ref{sec:CMOS}.   The   perspective    of   automating   speckle
interferometry    in   the    future   is    briefly   discussed    in
Section~\ref{sec:disc},    and   the    results   are    summarized   in
Section~\ref{sec:sum}.

%---------------------------------------------------------
\section{Performance of HRCam at SOAR}
\label{sec:perf}

The HRCam speckle camera and its  data processing have been covered by
\citet{HRCAM}. Here,  the main  instrument parameters are  recalled for
consistency, and its photometric calibration is given. In the following
Section, this information is used for comparison with simulations.

%---------------------------------------------------------
\subsection{Instrument Parameters}
\label{sec:det}

Since 2017, the EM CCD  detector iXon X3 888 from Andor\footnote{\url{
 www.andor.com}} has been in use  in HRCam.  The optical magnification
has been adjusted to project its 13 $\mu$m square pixels at a scale of
$p=   0\farcs01575$   per   pixel   on  the   sky  to   sample   the
diffraction-limited  speckles adequately.   For  a telescope  diameter
$D=4.1$\,m, the critical sampling of  $\lambda/(2 D)$ corresponds to a
pixel scale of $0.0135''$ and  $0.0207''$ at a wavelength $\lambda$ of
540  and 822  nm, respectively,  corresponding  to the  two most  used
spectral bands.   The speckle image  is slightly undersampled  in the
green filter $y$ and oversampled in the $I$ band by 1.3 times.

The iXon camera  has a conversion factor of $g  = 10.1$ electrons (e-)
per  Analog to  Digital Unit  (ADU), as  specified by  the vendor  and
confirmed by our measurements.  The  EM gain $E_g$ setting corresponds
to the actual signal amplification, to  within a few per cent. The rms
readout  noise (RON)  is 4.5  ADU  or 45  e-, and  its impact  becomes
negligibly small  at $E_g  > 100$.   A typical  histogram of  the dark
signal shows a Gaussian distribution with the width that characterizes
the RON and an exponential tail produced by the single-electron events
\citep[see Figure  3 in][]{HRCAM}.   These clock-induced  charge (CIC)
events  do not  depend  on  the exposure  time  (the  dark current  is
negligibly small  because the  CCD is  cooled to  $-60^\circ$\,C), and
their rate  is quantified  by the fractional  area of  the exponential
tail in the  dark-signal histogram; it is about 0.02  events per pixel
per frame.  In this camera, the CIC rate depends on the vertical clock
time, which has  been chosen to minimize the rate.   Parameters of the
detector,  remeasured in  2023, show  no degradation  relative to  our
measurements in 2016, despite its intensive use for 7 yr.

%---------------------------------------------------------
\subsection{Data Acquisition and Processing}
\label{sec:proc}

So far, the operation  of HRCam relies on the human  experience and on the
quick-look evaluation  of results  immediately after  acquisition. The
targets are selected flexibly,  depending on the observing conditions.
Most  HRCam   data  are  acquired   in  the  standard  mode,   with  a
200$\times$200 pixels  region of interest  (ROI), 400 images  per data
cube, without binning.  The field  of view, 3\farcs15, is large enough
to  capture seeing-limited  images  without  truncation.  The  minimum
exposure time  in this  mode is 24.4\,ms,  and the  effective exposure
time (interval between successive images) is 27.9\,ms.  Acquisition of
a standard data  cube thus takes 12\,s, and two  cubes per observation
are normally recorded with  an EM gain of $E_g =100$.  These parameters
are adopted in the simulations.  Faint targets are observed in the $I$
filter; its transmission and the detector spectral response define the
bandwidth with an average wavelength of 822 nm and a width of 140\,nm.
Note, however, that both parameters depend on the stellar temperature,
and the effective  response becomes ``redder'' for red  stars. The $y$
filter (543/22 nm) is used for  observations of bright and close binaries
with maximum angular resolution.

\begin{figure}[ht]
\epsscale{1.1}
\plotone{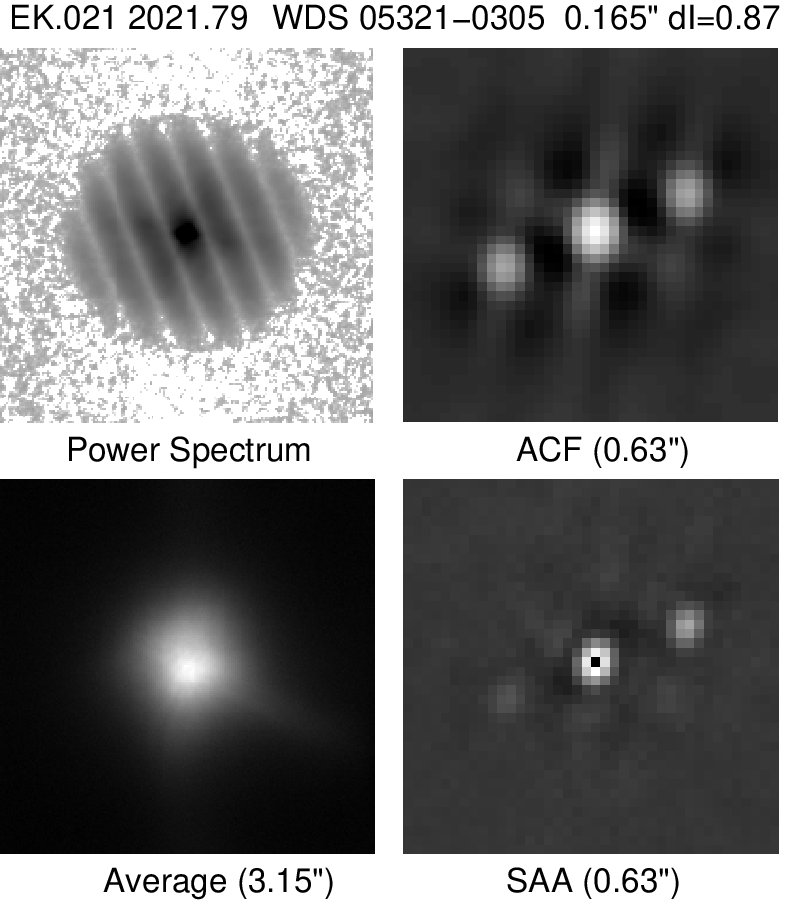}
%\plotone{Data.eps}
\caption{Example  of  processing  the  data cube  EK.021  recorded  on
  2021.79.  The  binary star WDS 05321$-$0305  (separation 0\farcs165,
  $\Delta I = 0.87$ mag) was  observed in the $I$ filter with standard
  parameters  (ROI 200$\times$200  pixels, exposure  time 24\,ms,  400
  frames per  cube). The speckle  PS is shown in  negative logarithmic
  stretch.  The  central 40$\times$40 pixels (0\farcs63)  fragments of
  the ACF and SAA images are  displayed. The FWHM of the average image
  is  0\farcs58.   Under  good  seeing, residual  aberrations  of  the
  telescope are manifested as asymmetric details of the average image,
  as a cross-like feature in the  PS, and as faint spurious details in the
  ACF.
\label{fig:data} 
}
\end{figure}

 The standard speckle interferometry \citep{Labeyrie1970} is based
  on  the evaluation  of the  {\em speckle  power spectrum}  (PS) $P$,
  which is the average square modulus of the Fourier transform (FT) of
  each image  $I_i$ in  the data  cube: 
\begin{equation}
P( {\mathbf f}) = \frac{1}{N_z} \sum_{i=1}^{N_z} | \tilde{I}_i ( {\mathbf f}) |^2 ,
\label{eq:pow}
\end{equation}
 where ${\mathbf f}$ is the spatial frequency, $N_z$ is the number
  of frames in the data cube,  and tilde denotes the FT operator. From
  each data cube,  the  speckle pipeline creates four two-dimensional
images: the PS, the speckle  autocorrelation function (ACF), which is
the inverse FT  of  the  PS,  the  shift-and-add  (SAA)  or  ``lucky''  image
registered  on  the  brightest  pixels, and  the  average  image  with
correction of the overall image centroid motion; see  details and
illustrations in \citet{HRCAM} and in Figure~\ref{fig:data}.

\begin{figure}[ht]
\epsscale{1.1}
\plotone{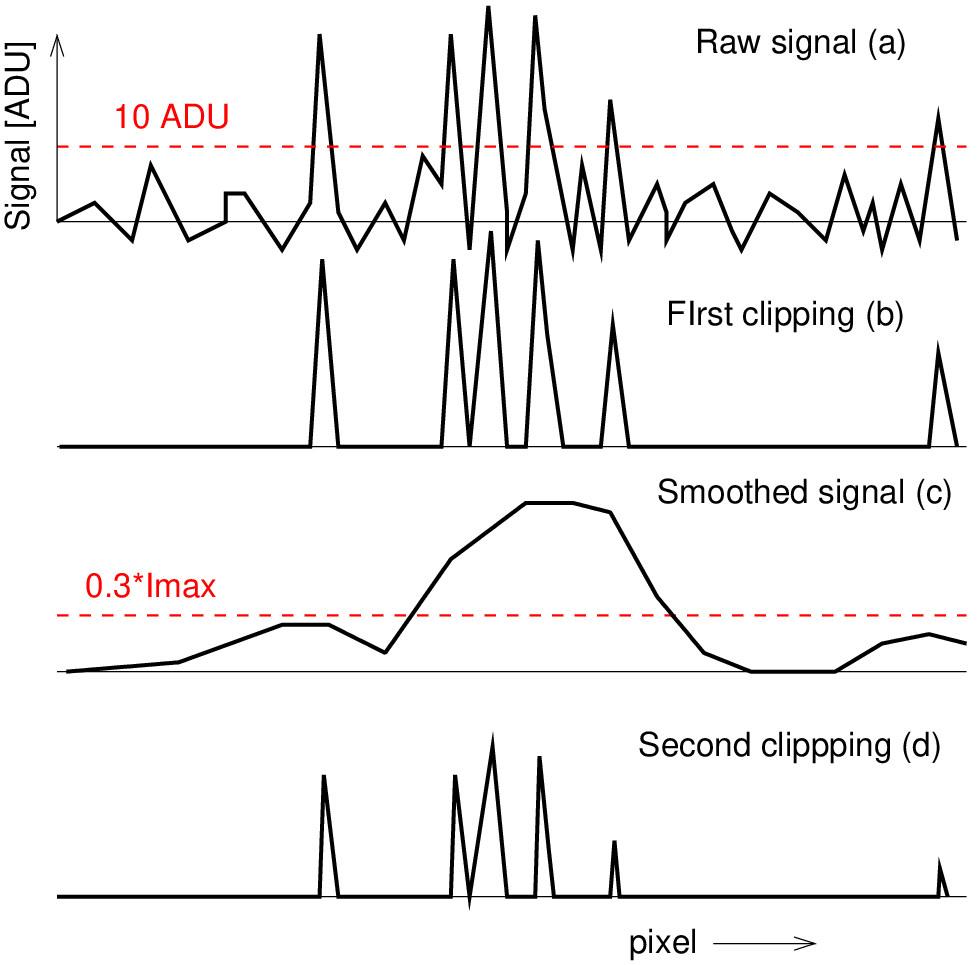}
%\plotone{Threshold.eps}
\caption{Illustration   of   the   standard   signal   processing   in
  HRCam.  Signal in  a  single  row of  the  bias-subtracted image  at
  various  processing stages  is  plotted by  thick  black lines,  the
  thresholds are shown in red; see  the text.
\label{fig:threshold} 
}
\end{figure}

The algorithm of  the PS calculation has been ``trained''  on the real
data to enhance the speckle signal  from faint sources.  It is illustrated
in Figure~\ref{fig:threshold}.  The average number  of CIC events in a
200$\times$200  frame,  800,  becomes  comparable  to  the  number  of
detected stellar photons $N_{\rm ph}$  for faint stars,  seriously
  degrading  the sensitivity.   After  subtraction of  the bias 
  (i.e.  the  average image taken without  illumination), most pixels
are  empty,  containing only  the  RON, while  some  pixels
contain the speckle and CIC signals  (a). To reduce the effect of RON,
the signal  is clipped at  the level of 10  ADU, and all  pixels below
this level  are set to zero  (b).  The first clipping  still transmits
the CIC  events, which  affect calculation of  the image  centroid and
flux  for faint  stars.  So,  the clipped  signal is  smoothed with  a
5-pixel  square kernel,  and the  second  threshold equal  to the  0.3
fraction of  its maximum is  defined (c).   It is subtracted  from the
smoothed  signal, negative  pixels are  set to  zero, and  the clipped
smoothed image  is used  for the centroid  calculation to  produce the
average tilt-corrected image.  The second threshold is also subtracted
from the clipped signal (b),  and the resulting nonnegative image (d)
is used to  calculate the PS and  SAA.  The sum of  the doubly clipped
image (d) is  less than the sum of the  original bias-subtracted image
(a) by a  large ($\sim$7) factor,  so the  fluxes estimated from  the
  clipped images are severely biased.

The  simulations described  below confirm  that the  current algorithm
produces near-optimum results for faint stars. However, calculation of
the PS  does not need  the second clipping, and  the image (b)  can be
used  instead.  It  is important,  however,  to subtract  from it  the
average background because  the PS is normalized by its  value at zero
spatial frequency $f$, which equals the square of the total flux.  For
correct evaluation of the speckle power, only the stellar flux must be
used for normalization,  so the average background  must be subtracted
from  the clipped  image  in  (b) before  calculation  of  the FT  and
summation  of  its  square  modulus  over  all  frames  in  the  cube.
Otherwise, the  CIC background  contributes to  the flux,  producing a
spike in  the PS at $f=0$.   The second clipping used  in the standard
algorithm  reduces  the background  effect  but  does not  cancel  it
completely.   So, the  original algorithm  of the  PS calculation  was
modified  to use  only  single clipping  with background  subtraction.
Other data  products (average and  SAA images) are still  derived from
the doubly clipped images.

%---------------------------------------------------------
\subsection{Photometric Calibration of HRCam}
\label{sec:ptm}
% see checkHRCam.tex

During 2022--2023,  a large number of  red dwarfs from the  Gaia
Catalog of Nearby  Stars (GCNS) were observed  with HRCam \citep{GKM}.
Parameters of  1325 such observations  (flux, exposure time,  EM gain,
and coordinates) were retrieved from  the general speckle database and
matched to  the GCNS sources to get their photometry  (magnitudes in
the Gaia $G$, $G_{BP}$, and $G_{RP}$  bands).  Fluxes  in the database,
strongly biased  by the image clipping (see above), were
replaced by fluxes measured on  the saved average images.

Given the  measured flux $F$ in  ADU per frame, the  conversion factor
$g=10.1$ e- per ADU, the exposure time $t$ in seconds, and the EM gain
$E_g$,  the  instrumental  magnitudes  of  HRCam  $m_{\rm  inst}$  are
computed as
\begin{equation}
m_{\rm inst} = 25 - 2.5 \log_{10}[gF/(t E_g)] .
\label{eq:minst}
\end{equation}
The offset of 25 mag is  arbitrary.  The spectral response of HRCam in
the  $I$ filter  is  ``redder'' compared  to  the Gaia  $G$  band, so  the
difference  between instrumental  magnitudes  and $G$  depends on  the
$G_{BP}-G_{RP}$  color  of   the  star.  This  empirical   dependence  can  be
approximated by the linear formula
\begin{equation}
m_{\rm inst} \approx G - 0.61 -  0.35 \; (G_{BP}-G_{RP}) .
\label{eq:color}
\end{equation}

Hence the photometric zero point  of HRCam in the instrumental $I_{\rm
  HRCam}$  system  is 25.6  mag,  and  such star  gives  a  flux of  1
el~s$^{-1}$.  The instrumental magnitudes  can be estimated as
\begin{equation}
I_{\rm  HRCam} \approx G -  0.35 (G_{BP}-G_{RP}). 
\label{eq:I}
\end{equation}
A  similar  comparison of  fluxes  in  the  $y$  filter with  the  $V$
magnitudes of observed stars results in the zero point of 24.6 mag. No
color term is necessary because the central wavelengths of the $y$ and
$V$ bands  are similar.  The  zero point in  the $y$ band  is brighter
than in $I_{\rm HRcam}$ by 1 mag  owing to the smaller bandpass of the
green filter.

%---------------------------------------------------------
\subsection{Measurement of Binary Stars}
\label{sec:bin}

Parameters  of   a  binary  star  (relative   position  and  magnitude
difference)  are  determined  by  fitting  a model  to  the  observed  PS
$P({\mathbf f})$ :
\begin{eqnarray}
P({\mathbf f}) & \approx & P_0({\mathbf f})[1 - B + B \cos (2 \pi {\mathbf f
  } {\mathbf x}) ],  \label{eq:fringe} \\
B & = & 2 r/(1 + r^2), \\ 
r & = & 10^{-0.4 \Delta m}, \label{eq:binary}
\end{eqnarray}
where ${\mathbf f}$  and ${\mathbf x}$ are  two-dimensional vectors of
the spatial  frequency and binary  position, respectively, $B$  is the
contrast of fringes in the PS  which depends on the intensity ratio of
the two components $r$ \citep{TMH10}.   The $P_0({\mathbf f})$ term is
the  PS of  a single  (reference) star.   A more  general formula  for
triples is  given in  \citet{HRCAM}.  The model  is fitted  at spatial
frequencies  between  $0.2 f_c$  and  $0.8  f_c$,   where  $f_c  =
  D/\lambda$ is  the cutoff frequency.  For  relatively wide binaries
with separation $\rho \gg \lambda/D$,  the reference PS is obtained by
angular  averaging of  the   observed PS.   Otherwise,  a PS  of
another  object observed  in  the same  conditions can  be  used as  a
reference  (if the  reference  object  is a  binary,  its fringes  are
removed by division).

The measurement errors are a quadratic sum of four effects:
\begin{enumerate}
\item
Random errors of the PS caused by the speckle noise, photon noise, and
detector (Section~\ref{sec:snr}). 

\item
Distortions of  the single-star PS  caused by optical  aberrations and
vibration, partially accounted for by using a reference star.

\item
Atmospheric errors caused by high-altitude turbulence \citep[see their
  estimate in][]{SAM21}.

\item
Inaccurate calibration of the pixel scale and orientation.

\end{enumerate}

The errors  (3) and  (4) increase with  the binary  separation $\rho$.
Analysis of  calibrator  binaries in \citet{SAM21} indicates that
$\sigma  \approx   0.81  +  1.15   \rho$  mas,  where  $\rho$   is  in
arcseconds. External  errors as small as  $\sim$1\,mas were documented
by the residuals  to good orbits of bright close  pairs; residuals are
larger for  binaries with a large  $\Delta m$.  The random  errors are
estimated  from the  formal errors    found by  fitting the  model
  (\ref{eq:fringe})  to  the observed  PS  and  from the  differences
between the results of two data cubes (whichever is larger).  They are
given in  the data  tables, and  a typical  median value  is 0.3\,mas.
Random  errors  dominate   for  faint  stars;  they   are  studied  in
Section~\ref{sec:err}.

%---------------------------------------------------------
\section{Simulations}
\label{sec:simul}

Simulation of  speckle data is  implemented in IDL (Appendix).   It is
split   into   two  parts.    First,   a   noiseless  data   cube   is
generated. Then,  the effect of  all noise sources is  simulated. This
allows  us   to  study  the   noise  using  the  same   input  speckle
pattern. Real  data on a  single bright star can  be used as  well for
simulating  the noise  and optimizing  the data  processing algorithm.
The theory  of light propagation  through the atmosphere  and relevant
parameters such as  seeing, Fried radius $r_0$,  and speckle coherence
time   $\tau$   are   covered  in,   e.g.,   \citet{Roddier1981}   and
\citet{Goodman1985}.  Most  simulations adopt a seeing  of 0\farcs8 at
800\,nm which corresponds to $D/r_0  = 20$.  The temporal evolution of the
speckles is modeled by two phase screens of equal strength moving with
the speeds of 8 m~s$^{-1}$ and 40 m~s$^{-1}$ in orthogonal directions.

%---------------------------------------------------------
\subsection{Effects of Exposure Time and Spectral Bandwidth}
\label{sec:speckle}
% see HRCamsimul.tex

\begin{figure}[ht]
\epsscale{1.1}
\plotone{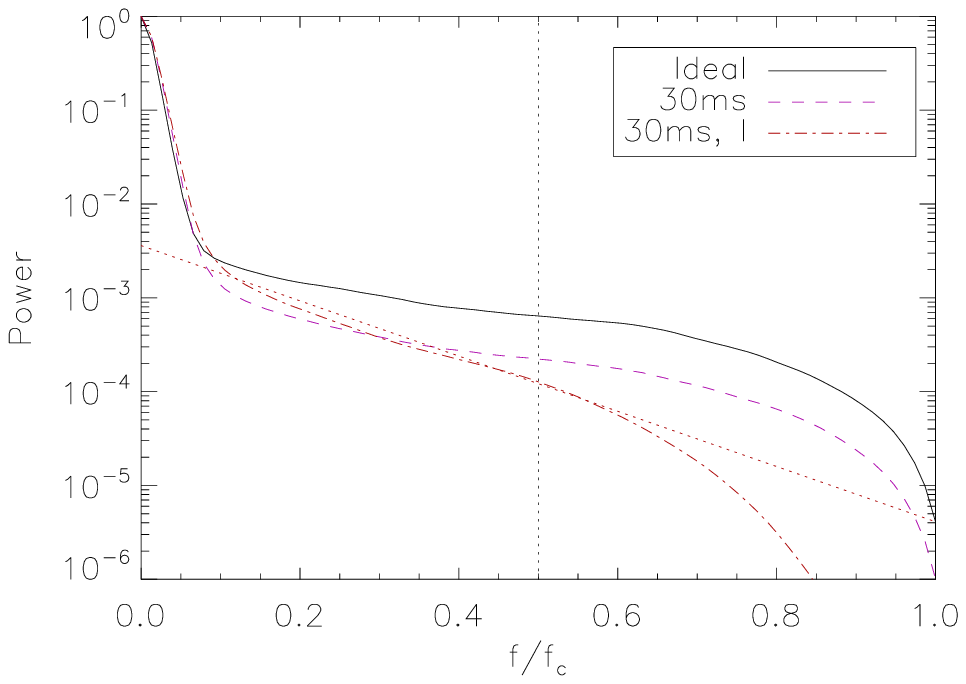}
%\plotone{bandwidth-test5.ps}
\caption{Azimuthally averaged PS of  simulated noiseless speckle cubes
  for a  0\farcs8 seeing:  an ideal (instantaneous  and monochromatic)
  speckle  at  800\,nm, a  monochromatic  exposure  of 30\,ms  (dashed
  line), and imaging in the  $I$ band (dash-dotted). The vertical dotted line
  marks $0.5  f_c$, the red dotted  line is a linear  approximation of
  the $I$-band PS.
\label{fig:bandwidth} 
} 
\end{figure}
% speckle/simul/test5.pro

The PS of  speckle images has two components:  the seeing-limited core
at low spatial frequencies and the high-frequency part extending up to
the cutoff frequency $f_c =  D/\lambda$.  Only the high-frequency part
corresponding to  the speckle signal  is of  interest here. The  PS is
normalized   to   one   at   zero  frequency;   it   is   rotationally
symmetric. According  to the theory,  in an ideal speckle  pattern the
high-frequency part of the PS  $P(f)$ should be approximately equal to
$0.435  (r_0/D)^2 T_0(f)$,  where $r_0$  is  the Fried  radius at  the
imaging wavelength  and $T_0(f)$  is the  diffraction-limited transfer
function   \citep{Roddier1981,Goodman1985};  $T_0(0.5)=   0.39$.   For
$D/r_0 = 20$, we get $P = 4.2 \times 10^{-3}$ at $f = 0.5 f_c$.  It is
convenient to use logarithmic quantities,  so the level of the speckle
signal  (also called speckle  contrast) is characterized  by the
parameter $S =  \log_{10} P(0.5f_c)$; in the  above numerical example,
$S = -3.37$.

Reduction of the speckle contrast due to finite exposure time and wide
spectral     bandwidth     is     well    understood     in     theory
\citep{Roddier1981,Tok1980}    and     confirmed    by    measurements
\citep{Karo1978}.  The characteristic time  of the speckle ``boiling''
$\tau =  r_0/\Delta V$ depends  on the turbulence-weighted  wind speed
dispersion $\Delta V$  and is typically a few  milliseconds at visible
wavelengths. When the  exposure time $t$ is  significantly longer than
$\tau$, the  PS is reduced  by a factor  of $\tau/t$ uniformly  at all
spatial  frequencies. A  finite  spectral  bandwidth $\Delta  \lambda$
causes radial elongation  of the speckle structure  reaching $ (\Delta
\lambda/\lambda) \; (\lambda/r_0)$ at the  periphery of the image (its
radius  is  $\lambda/r_0$).  To  preserve  the  speckle contrast,  the
elongation must be  less than the speckle size  $\lambda/D$, hence the
spectral  bandwidth is  limited to  $\Delta \lambda/\lambda  < r_0/D$,
e.g.   40\,nm  at $\lambda  =  800$\,nm  and  $D/r_0  = 20$.   As  the
bandwidth of the  HRCam $I$ filter is wider,  the chromatic elongation
is  nonnegligible,  and $P(f)$  is  reduced  mostly at  high  spatial
frequencies.

Figure~\ref{fig:bandwidth} shows how the  speckle signal is reduced by
finite exposure time and wide  spectral bandwidth.  The same seeing of
0\farcs8  was  simulated using  three  different  codes of  increasing
complexity  (see Appendix).   The $\log_{10}P(f/f_c)$  of the  real
speckle data is approximated by a  straight line between $0.2 f_c$ and
$0.8 f_c$, and these parameters  (intercept $p_0$ and slope $p_1$) are
determined for the simulated data  as well.  Obviously, $S \approx p_0
+  0.5   p_1$.   Table~\ref{tab:contrast}  lists  the   PS  parameters
corresponding to  Figure~\ref{fig:bandwidth}.  The Full Width  at Half
Maximum (FWHM)  of the simulated  average centered images is  given in
the  last  column  in  arcseconds.    It  is  consistent  between  the
simulations and less  than the input seeing because  the overall image
motion has  been compensated  by centering.  We  note that  the finite
exposure time  degrades the speckle  contrast almost uniformly  at all
frequencies, while the slope $p_1$  becomes only slightly steeper; the
PS shape  resembles $T_0(f)$, as  in the monochromatic  case.  Uniform
reduction  by a  factor of  3 at  30 ms  exposure implies  a speckle
lifetime of $\tau \approx 10$\,ms.  On  the other hand, the wide spectral
bandwidth degrades mostly the high-frequency speckle power, and the PS
becomes steeper compared to the monochromatic light case.

\begin{table}
\center
\caption{Speckle Signal in Simulations}
\label{tab:contrast}
\medskip
\begin{tabular}{l c  c  c c} 
\hline
Case  & $S$  & $p_0$ & $p_1$ & FWHM \\
\hline
Ideal       &   $-$3.20 &$-$2.81 &$-$0.78 &  0.69 \\
30\,ms      &   $-$3.67 &$-$3.19 &$-$0.92 &  0.63 \\
30\,ms, $I$ &   $-$4.03 &$-$2.44 &$-$2.94 &  0.61 \\
\hline
\end{tabular}
\end{table}

\begin{figure}[ht]
\epsscale{1.2}
\plotone{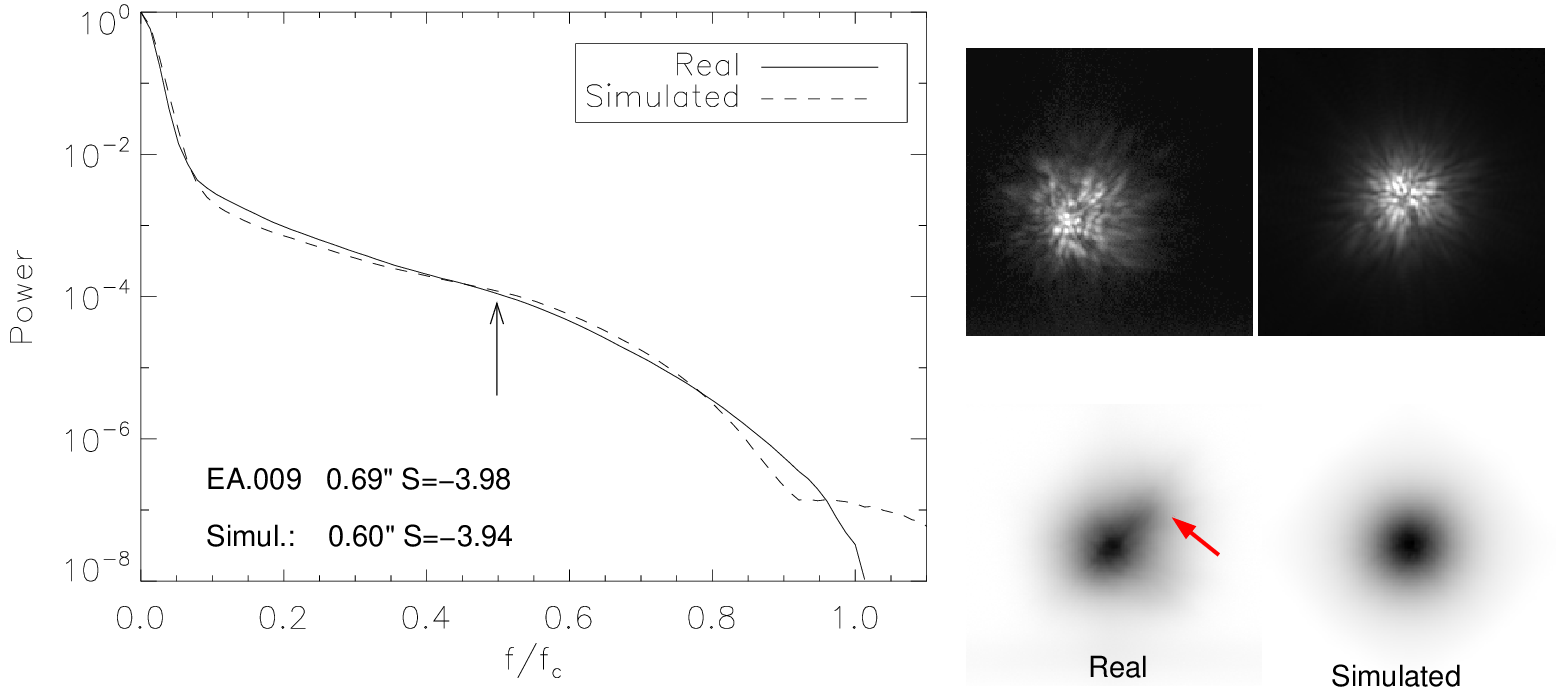}
%\plotone{Testsimul.eps}
\caption{Comparison between real and simulated data in the $I$ filter.
  Left: PS  averaged in angular coordinate;  right: individual speckle
  images (top) and the average  images in negative rendering (bottom).
  Note that the linear fit to PS between  0.2 and 0.8 $f_c$ (not plotted) is
  a poor approximation in this case.
\label{fig:poly} 
}
\end{figure}

For comparison with simulations, a  real image cube EA.009 recorded on
January 28, 2021 was selected.  This corresponds to a bright reference
star observed  in the $I$  filter with  an effective exposure  time of
27.8\,ms.  The  seeing was relatively  good, and the average  image is
almost round with a FWHM of 0\farcs69.  Figure~\ref{fig:poly} compares
the real  and simulated speckle  PS.  The  values of $S$  match almost
exactly.  The  FWHM of the  average (centered and  coadded) simulated
image  is  0\farcs61; it  is  smaller  than  the simulated  seeing  of
0\farcs80 owing to the centering  (tilt compensation).  The real image
is slightly larger, but it shows signs of residual aberrations (note a
small ``tail'' marked by the red arrow).

\begin{figure}[ht]
\epsscale{1.1}
\plotone{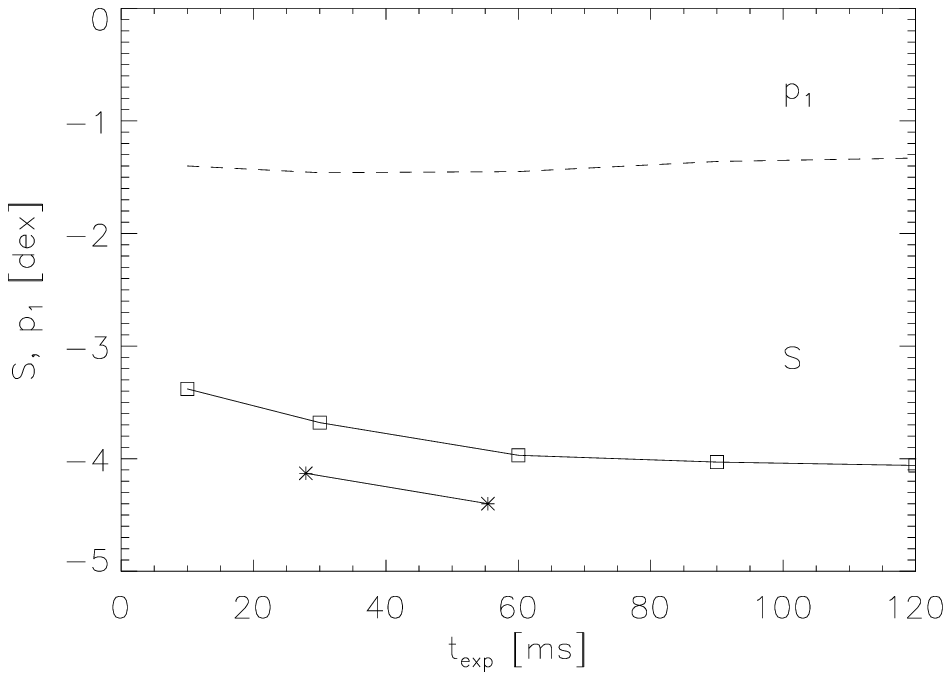}
%\plotone{simul-exptime.ps}
\caption{Dependence of the speckle signal $S$ (solid line and squares)
  and the PS slope $p_1$ (dashed  line) on the exposure time according
  to  simulations. Asterisks  show  results for  the  real image  cube
  EA.009.
\label{fig:exptime} 
}
\end{figure}

Simulations  help  us to  optimize  the  exposure time  of  individual
frames. With increasing exposure, a  larger number of photons from the
star is  collected, but at the  same time, the speckle  signal $S$ is
reduced.   The  effect of  increasing  exposure  time is  explored  in
Figure~\ref{fig:exptime}.   An exposure  increase  from  30 to  60\,ms
reduces the  speckle signal $S$  by 0.3\,dex  (by a factor of  2). This
reduction is almost  uniform at all spatial frequencies,  so the slope
$p_1$  remains   approximately  constant  (see  the   dashed  line  in
Figure~\ref{fig:bandwidth}).   Asterisks  show  results for  the  real
image  cube EA.009  and for  the  cube with  frames binned  (averaged)
pairwise.  Binning of  frames reduces the speckle  signal from $-4.13$
to $-4.40$ dex, also by a factor of 2. 

\begin{figure}[ht]
\epsscale{1.1}
\plotone{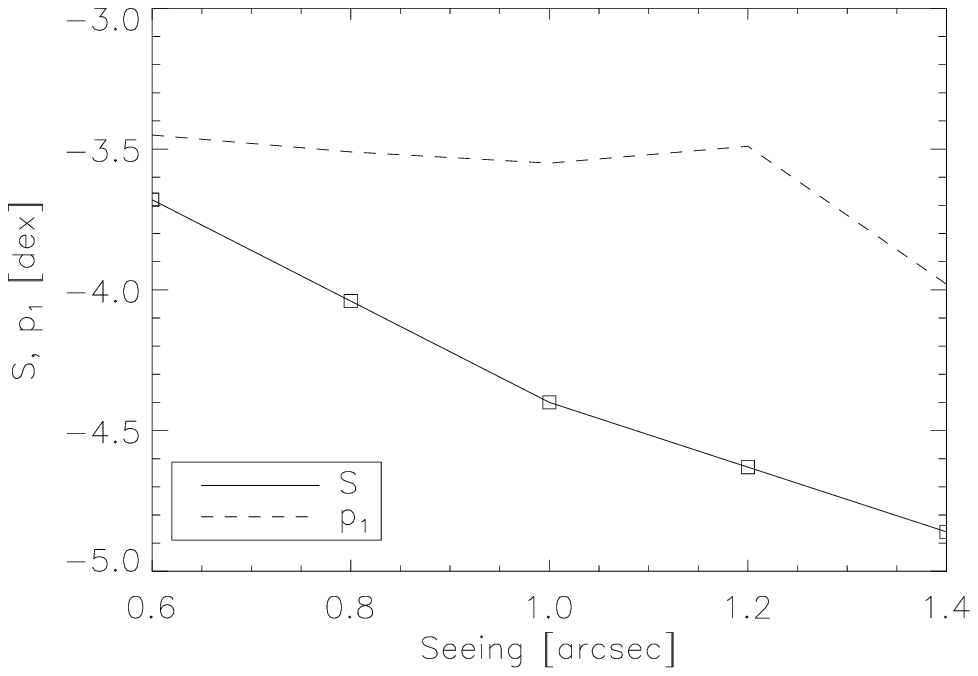}
%\plotone{s-seeing.ps}
\caption{Dependence of the speckle signal $S$ (solid line and squares)
  and the PS slope $p_1$ (dashed  line) on seeing.  Simulations with a
  30\,ms exposure time and the $I$ filter.
\label{fig:s} 
} 
\end{figure}

Figure~\ref{fig:s} shows  how the level  of the speckle signal  in the
$I$ filter  degrades with  increased seeing.  The  slope $p_1$ also becomes steeper, mostly owing to the  finite spectral bandwidth.

\begin{figure}[ht]
\epsscale{1.1}
\plotone{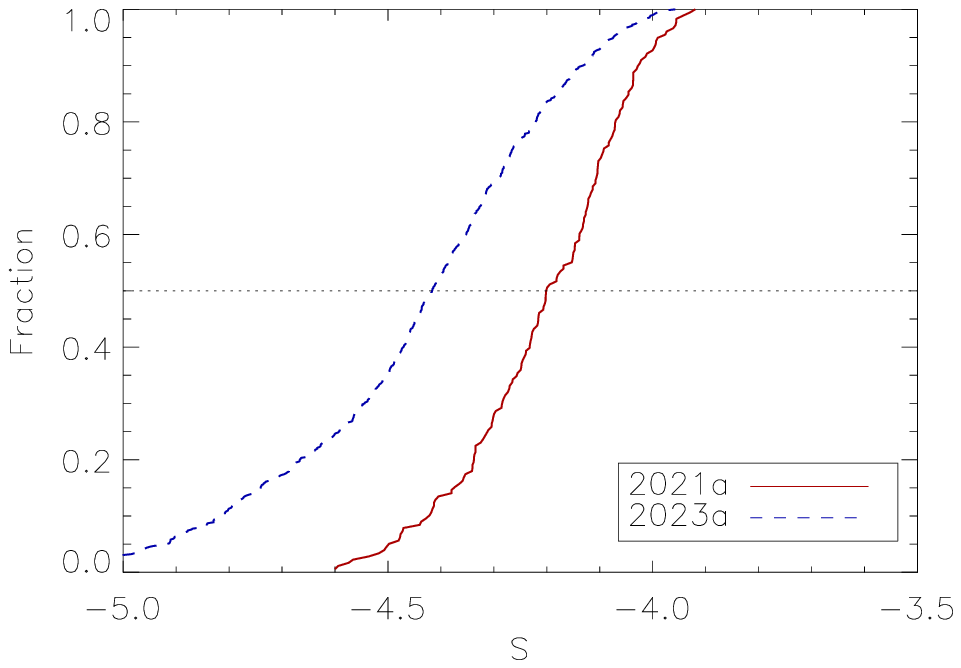}
%\plotone{S2123a.ps}
\caption{Cumulative histograms  of the speckle  signal $S$ in  the $I$
  filter without binning  in the speckle runs of 2021a  and 2023a with
  good and  average seeing, respectively.   The median $S$  values are
  $-4.19$  and $-4.42$ and  the median  PS slopes  $p_1$ are  $-3.78$ and
  $-3.61$.
\label{fig:shist} 
}
\end{figure}

%s21: 178  median  -4.19  median p1 -3.78
%s23: 620  median -4.42  median p1 -3.61

Cumulative  histograms  of  the  actual  speckle  signal  $S$  in  two
observing  runs are  shown in  Figure~\ref{fig:shist}.  Note  that the
standard processing algorithm with double clipping slightly biases $S$
to  smaller values,  and  this  bias depends  on   the number  of
  detected  photons  per  frame  $N_{\rm  ph}$,  as  demonstrated  by
processing  simulated data  with single  and double  clipping.  Double
clipping has little effect on bright stars.

Compared to the  simulations, the speckle signal in the  real data may
be  reduced   by  additional  factors  such   as  optical  aberrations
(e.g. imperfect focusing) and telescope vibration. The reasonably good
match  between real  data and  simulations is  encouraging, suggesting
that  these  additional degrading  factors  play  a minor  role.   The
atmospheric  parameters  adopted in  the  simulations  (e.g. the  wind
speed)  differ  from  the actual  (unknown)  parameters.   Approximate
agreement with  the real data  justifies the use of  these simulations
for evaluation of the HRCam performance.

%---------------------------------------------------------
\subsection{Signa-to-noise Ratio and Limiting Magnitude}
\label{sec:snr}

The noise simulator reads a simulated noiseless image cube from a fits
file, generates pixel values affected by the detector noise and random
numbers  of photons,  computes  the PS,  and  determines its  relevant
parameters. The  amplification noise  in an EM  CCD is  also simulated:
each  photon  generates a  random  signal  distributed as  a  negative
exponent with  a decrement $A$  which equals the average  amplitude of
single-photon events, and randomly amplified signals of all photons in
a pixel  are summed up.  The photon numbers  in each pixel  follow the
Poisson distribution  with an average  value equal  to the sum  of the
stellar photons and CIC. Simulating the HRCam detector with $E_g=100$,
I adopt $A=10$ and a CIC rate of  0.02.  A threshold of 10 ADU is used
in the PS calculation.

If a perfect detector records $N_{\rm  ph}$ photon events per frame on
average, the  PS increases  by $P_{\rm  bias} =  1/N_{\rm ph}$  at all
frequencies owing to the  photon noise bias \citep{Goodman1985}.  This
theoretical result has  been reproduced in the  simulation.  The value
of $P_{\rm  bias}$ is evaluated  by averaging the PS  over frequencies
beyond $f_c$, where the speckle power  is zero. The bias is subtracted
from  the  PS  before   its  azimuthal  averaging  and  model fitting.
Additional noise sources increase the  PS bias, reducing the effective
number of photons,  and this reduction factor  $1/(P_{\rm bias} N_{\rm
  ph})$   is  a   convenient  dimensionless   characteristic  of   the
sensitivity  loss caused  by a  noisy detector  in comparison  with an
ideal one.

For  bright stars,  the  photon noise  should  dominate over the detector
noise, and  the relation  $P_{\rm bias} =  1/N_{\rm ph}$  should hold.
So, the  measured values  of $P_{\rm  bias}$ can  be used  to estimate
$N_{\rm ph}$  and, by comparing  it with  the recorded flux,  to check
$g$, the  conversion factor  of the detector.   Analysis of  the HRCam
data  revealed that  the flux  recorded in  the database  was strongly
reduced   by  double   clipping   (see  above).    With  the   correct
(recomputed)  fluxes, the  expected linear  relation between  $P_{\rm
  bias}$ and $1/N_{\rm ph}$ is  retrieved, leading consistently to the
gain  factor $g  = 4.7$  e-/ADU, 2  times less  than measured.   The
reason for  this apparent discrepancy  is the amplification  noise.  It
doubles the signal variance, compared  to a pure Poisson distribution,
and  effectively halves  $N_{\rm ph}$.   Simulation of  an EM  CCD with
amplification noise indeed  shows that $P_{\rm bias}  = 2/N_{\rm ph}$.
So, the HRCam data are consistent with simulations.  For bright stars,
where the  speckle noise  dominates anyway,  the SNR  loss due  to the
amplification  noise is  not  detrimental, but  it  matters for  faint
stars, as well as CIC.

For one frame,  the variance of the FT square  modulus at each spatial
frequency equals the square of its mean  value (this is a  consequence of
the   negative-exponential  distribution   of  the   square  modulus).
Averaged over  $N_z$ frames, the  relative rms fluctuations of  the PS
equal $1/\sqrt{N_z}$  \citep{Dainty1974}. Let  $10^S$ be the  value of
the speckle power at half of the cutoff frequency, identified with the
useful signal.  The  total PS signal is $10^S +  P_{\rm bias}$, so the
SNR in one element of the PS is
\begin{equation}
SNR = \sqrt{N_z} \; 10^S/(10^S + P_{\rm bias}) .
\label{eq:SNR}
\end{equation}
For  bright  stars,  $10^S  \gg   P_{\rm  bias}$,  the  speckle  noise
dominates, and the SNR tends  to 20  for $N_z =  400$.  For  faint stars,
$P_{\rm bias}$ becomes the dominant term  in the PS and determines the
SNR.   The transition  between  the  bright-star  and  faint-star  regimes
depends on the  level of the speckle signal $S$,  it is around $N_{\rm
  ph} \sim 10^4$ for $S = -4$.

\begin{figure}[ht]
\epsscale{1.2}
\plotone{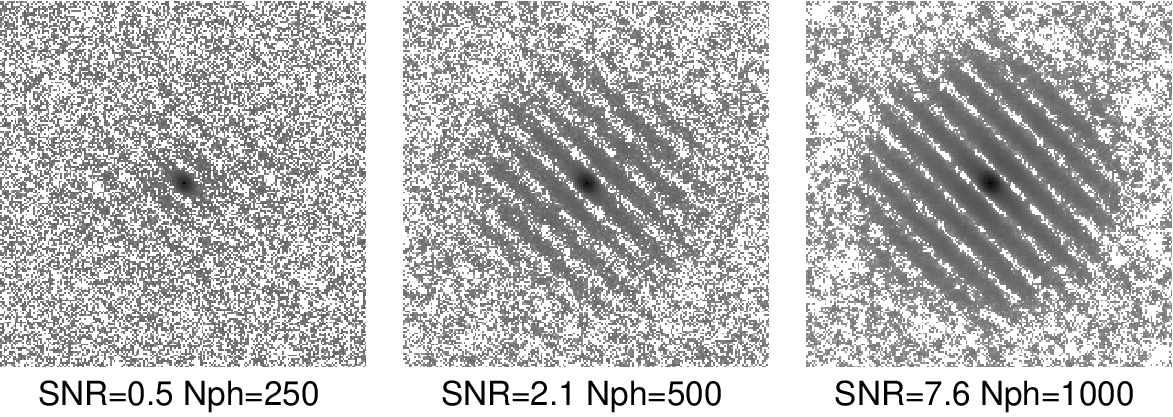}
%\plotone{Binaries.eps}
\caption{Power  spectra   of  a  simulated  binary   star  with  equal
  components displayed on  a negative logarithmic scale.   The SNR and
  $N_{\rm ph}$  are indicated.  An EM CCD  with CIC=0.02,  $A=10$, and
  threshold of  10 ADU is  simulated. An  ideal speckle data  cube was
  used as input.
\label{fig:binary} 
}
\end{figure}

Optionally, a binary star can be simulated by adding a shifted copy of
the  speckle  pattern scaled  by  the  binary intensity  ratio  before
simulating the noise.  Figure~\ref{fig:binary} shows fringes in the PS
of  a binary  star  at different  photon fluxes.   Note  that the  SNR
parameter  here refers  to single  stars (fringes  reduce the  average
level of  the PS and  further decrease the SNR).   At an SNR of  0.5, the
fringes are almost lost in  the fluctuations of the photon noise bias,
but    the   binary    can    still   be    detected   and    measured
(Section~\ref{sec:err}).

\begin{figure}[ht]
\epsscale{1.1}
\plotone{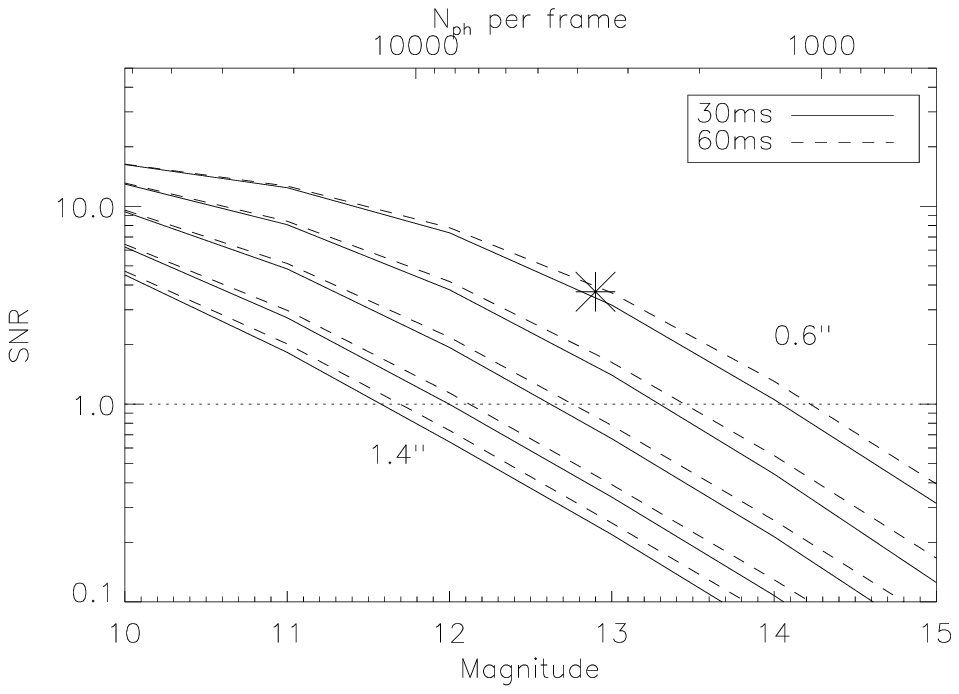}
%\plotone{snr-test4a.ps}
\caption{Signal-to-noise ratio in  the simulated HRCam with  an EM CCD
  detector vs. stellar magnitude in  the instrumental system $I$. Full
  lines and upper  axis correspond to the 30\,ms  exposure, and dashed
  lines correspond  to the 60\,ms  exposure, both with 400  frames per
  data cube.  The curves are computed  for seeing values of  0.6, 0.8,
  1.0, 1.2, and 1.4 arcsecond.   The asterisk shows the observation of
  V1311 Ori D in 2021.7983 (see text).
\label{fig:snr} 
} 
\end{figure}
% test4a.pro

With realistic  simulations of the  noisy speckle data, I  explore the
combined effect of the seeing variation and the source flux on the SNR
(Figure~\ref{fig:snr}). The instrumental $I$  magnitudes of stars that
reach SNR=1  range from  11.5 to  14.5 mag,  depending on  the seeing.
These  estimates  are slightly  optimistic  because,  compared to  the
simulation, the speckle contrast  is additionally reduced by imperfect
optics (e.g. a focus error) and  by telescope vibration.  On the other
hand, a slower wind in the upper atmosphere would increase $S$ and the
magnitude limit  relative to  the simulations.  Doubling  the exposure
time from 30 to  60 ms doubles the flux $N_{\rm  ph}$, but reduces the
speckle power  by the same amount  (in agreement with the  theory), so
the small net  SNR gain is due  to  secondary factors  like CIC; it
comes at the cost of doubling the acquisition time.

As a real example, consider observation  of the faint single red dwarf
V1311~Ori~D reported  in \citet{V1311Ori}.   On that  date, 2021.7983,
the seeing was very good. The FWHM  of the centered image in the data
cube EK.026 (exposure time 50\,ms) is 0\farcs63, the estimated speckle
signal  is  $S=-3.7$, and  the  SNR  is  3.7.  With  the  instrumental
magnitude of  12.92, estimated  from the $G=13.92$  mag and  the color
$G_{BP}-G_{RP}  \approx 2.4$  mag, the  experimental point  (asterisk)
falls near  the upper  curves in  Figure~\ref{fig:snr}. Note  that for
very  red  stars, the  effective  spectral  response differs  from  the
response assumed in the simulation.

%---------------------------------------------------------
\subsection{Dependence of Measurement Errors on the SNR}
\label{sec:err}

\begin{figure}[ht]
\epsscale{1.1}
\plotone{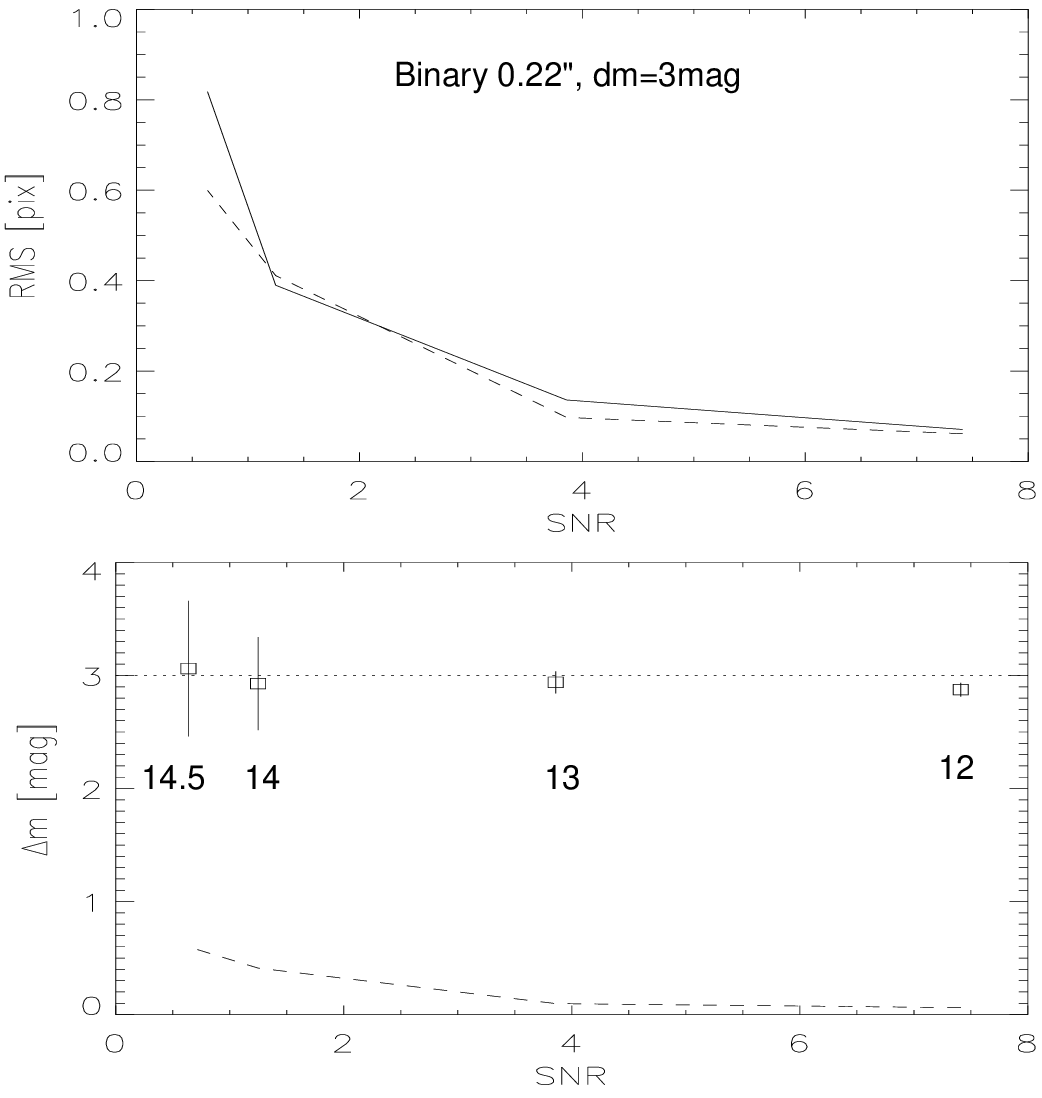}
%\plotone{Errors.eps}
\caption{Errors  in the  position  (top: the  solid  and dashed  lines
  correspond to  two orthogonal  directions) and  magnitude difference
  (bottom)  of simulated  binary  star measurements  depending on  the
  SNR. Approximate magnitudes  are indicated in the  lower plot, where
  the dashed  line shows  the errors, while  squares with  error bars
  show the mean fitted $\Delta m$ and their scatter.
\label{fig:errors} 
} 
\end{figure}
% testmeas.pro

Measurement  of the  binary stars'  parameters (relative  position and
magnitude difference) depends  on the random and  systematic errors of
the PS.   Most observed  binary stars are  bright, and  the systematic
errors  dominate  (Section~\ref{sec:bin}).   The influence  of  systematic
errors is reduced  by observations of a reference star  under the same
conditions, provided that the  instrument-related PS distortions
are stable.  An  example of erroneous measurement  caused by vibration
is given in Figure~6 in \citet{HRCAM}.

For faint stars, random errors dominate, and they can be quantified by
simulations. I simulated a binary  star with a separation of 0\farcs22
and a magnitude difference $\Delta m  = 3$ mag with combined magnitude
ranging  from 12  to  14.5  mag, observed  with  HRCam under  standard
conditions in  the $I$  filter.  For each  magnitude, 20  random noisy
cubes were  generated, and a  binary star model  was fitted to  the PS.
Figure~\ref{fig:errors} plots the results.  The SNR varies between 0.6
and 7.5. The relative flux of  the companion is only 0.063, and the fringe
contrast in the PS is 0.126. Despite this, the model fits to the PS do
not diverge even  at an SNR of 0.6  (remember that SNR is  defined at one
spatial  frequency,  while  the  parameter fit  uses  all  frequencies
between $0.2f_c$ and $0.8 f_c$).  The position errors reach 0.7 pixels
or 11\,mas. 

%---------------------------------------------------------
\section{Comparison between EM CCD and CMOS Detectors}
\label{sec:CMOS}

\begin{figure}[ht]
\epsscale{1.1}
\plotone{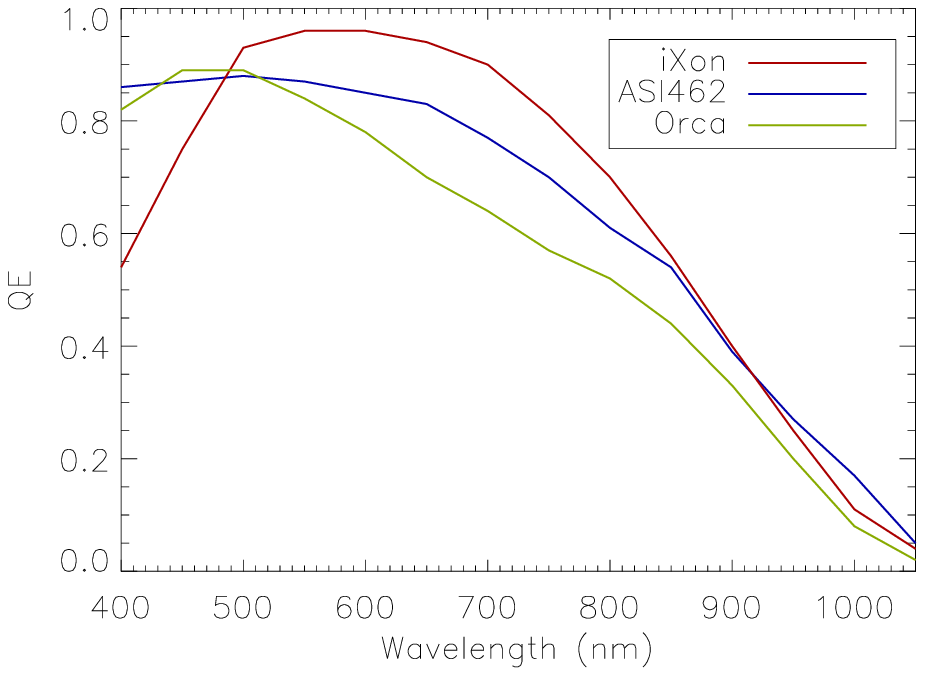}
%\plotone{qeplot.eps}
\caption{Quantum efficiency of three detectors for speckle
  interferometry according to the vendor's data: EM CCD iXon X3  888
  from Andor, ASI462 from ZWO, and ORCA-Quest from Hamamatsu.
\label{fig:qe} 
}
\end{figure}

Recently, the noise level of CMOS  detectors has improved to the point
where they have  become competitive with EM CCDs.   CMOS detectors for
astrophotography,             e.g.             ASI462             from
ZWO,\footnote{https://www.zwoastro.com/product/asi462mm/}   are  very
cheap  and  readily  available,  while  their RON  is  about  0.5  e-.
Hamamatsu  developed  a scientific  qCMOS  camera  ORCA Quest  with  a
subelectron
RON.\footnote{https://www.hamamatsu.com/us/en/product/cameras.html}
Its  indicative  cost  is  \$50K.   This  Hamamatsu  camera  has  been
installed in  the speckle  interferometer of the  2.5 m  telescope and
demonstrated     an     improvement      in     limiting     magnitude
\citep{Strakhov2023}.  A  CMOS  detector,  even with  a  0.5  e-  RON,
eliminates such  problems of EM  CCDs as CIC and  amplification noise.
The  quantum  efficiency  of   these  three  detectors  is  comparable
(Figure~\ref{fig:qe}).

\begin{figure}[ht]
\epsscale{1.1}
\plotone{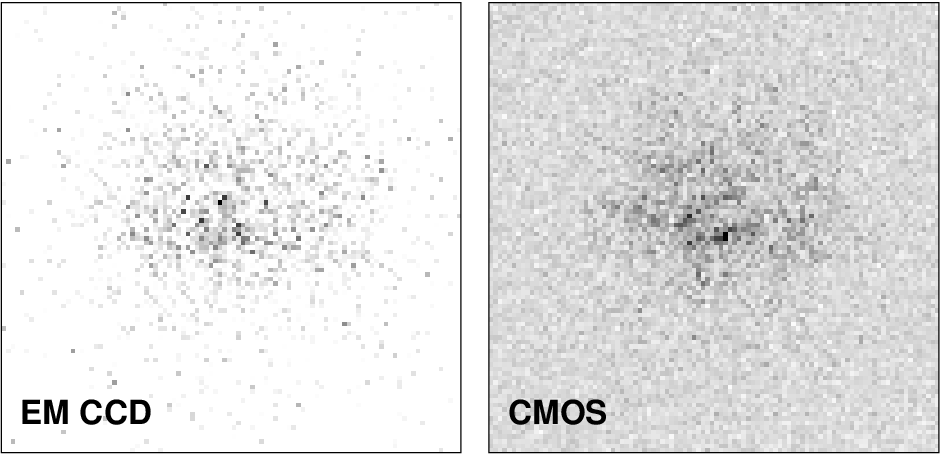}
%\plotone{Compare.eps}
\caption{Simulated   single  noisy   frames  with   $N_{\rm  ph}=3000$
  corresponding to an  EM CCD (left, SNR=1.7), and a  CMOS with RON=0.5
  e- (right, SNR=3.3).  The central region of 100$\times$100 pixels is
  shown in negative linear stretch.
\label{fig:frames} 
}
\end{figure}

These  detectors are  compared below  by additional  simulations.  The
amplification noise and  CIC are both absent for CMOS.  I assume a RON
of 0.5 e-, i.e.  a cheap CMOS  camera.  The EM CCD parameters are the same
as above.   Figure~\ref{fig:frames} compares simulated  speckle images
of a faint  source with the same $N_{\rm ph}  = 3000$ corresponding to
these detectors ($I$ filter, 30\,ms exposure time, $S = -4$ dex).

\begin{figure}[ht]
\epsscale{1.1}
\plotone{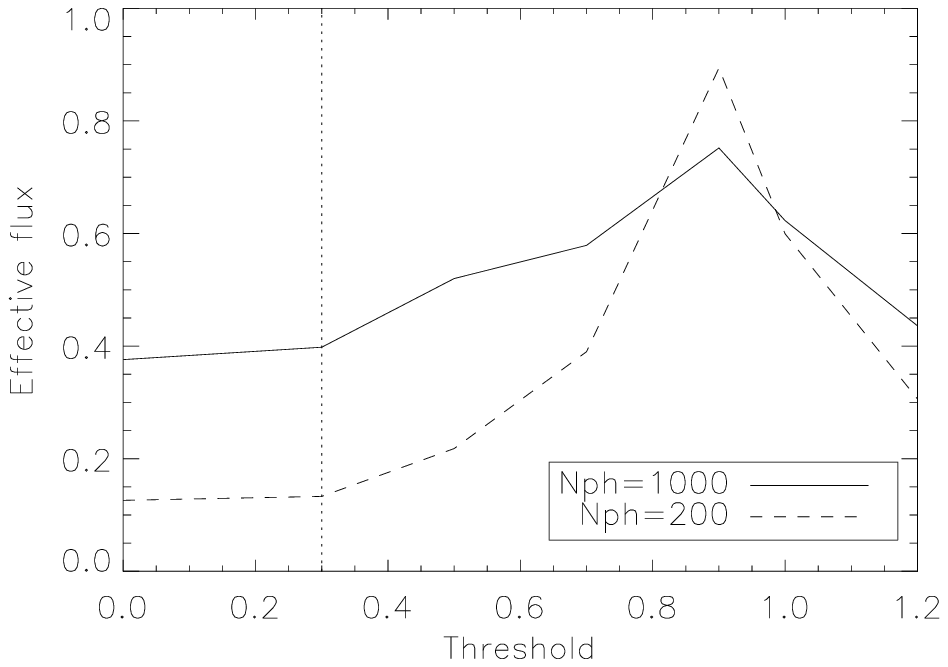}
%\plotone{threshold-new.ps}
\caption{Effective loss  of flux  $1/(P_{\rm bias} N_{\rm ph})$  vs. threshold
  for a  qCMOS with RON of 0.3 e- (marked  by the vertical dotted  line) and
  for two levels of the flux.
\label{fig:threshold2} 
} 
\end{figure}

Signal clipping in the  calculation of the PS is needed  for a CMOS to
reduce  the impact  of RON  in ``empty''  pixels that  do not  contain
stellar photons.  Figure~\ref{fig:threshold2} shows the effective flux
loss vs.   threshold for a  low-noise qCMOS  camera and two  levels of
photon flux.  The  optimum threshold appears to  be near 3$\times$RON.
For  a CMOS  with a RON  of  0.5 e-,  the same  $3\sigma$ threshold  is
adopted.    It   seems   that   clipping  has   not   been   used   by
\citet{Strakhov2023},  who found  a smaller  gain in  sensitivity when
switching from an EM CCD to a qCMOS.  Note also that their EM CCD has a
higher level of CIC events compared to HRCam (0.04 vs. 0.02).

\begin{figure}[ht]
\epsscale{1.1}
\plotone{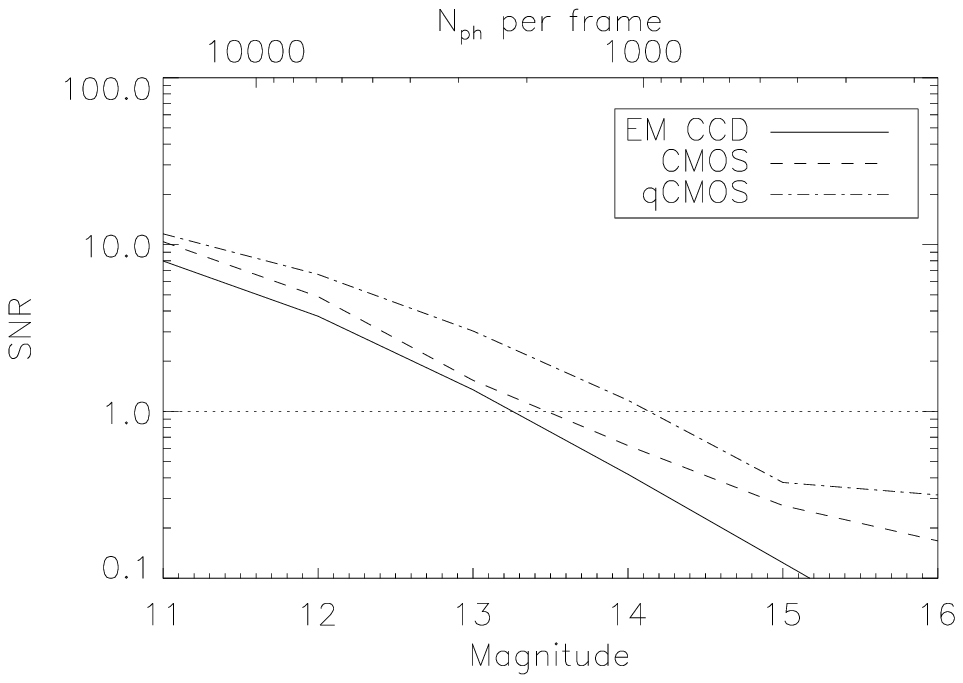}
%\plotone{snrplot2-new.ps}
\caption{Signal-to-noise  ratio (SNR) in the PS at  half of the cutoff
  frequency vs.   magnitude in the $I$  band for a 0\farcs8  seeing and
  three simulated  detectors: an EM CCD,  a scientific CMOS (RON  of 0.5
  e-), and a qCMOS  (RON of 0.3 e-).  The upper  axis gives the number
  of  photons per  frame in  a 30\,ms  exposure; the  SNR assumes  400
  frames per data cube.
\label{fig:snr2} 
} 
\end{figure}
% test4.pro

Figure~\ref{fig:snr2}  compares the  SNR vs.  magnitude for  the three
detectors. It assumes equal $N_{\rm ph}$ and an optimized threshold in
the PS calculation.  A realistic level  of the speckle signal $S = -4$
dex corresponding  to a 0\farcs8 seeing  and an $I$ filter  is adopted
(see Figure~\ref{fig:s}).  An upgrade from an  EM CCD to a qCMOS offers
a sensitivity gain of $\sim$1 mag.   Note that the SNR is computed for
$N_z = 400$.  Increasing the number  of acquired frames by a factor of
10 pushes  the curves  up by a  factor of 3,  and 16th  mag stars
become accessible with a qCMOS.

The SNR  plot in Figure~\ref{fig:snr2}  refers to the speckle  signal at
$0.5 f_c$.   In a  wide spectral  band like  $I$, the  PS has  a steep
slope, and  its value  at $0.2 f_c$  is larger by  almost an  order of
magnitude, $S  \sim -3$  dex (Figure~\ref{fig:bandwidth}).   So, speckle
observations  of  faint  stars  are  feasible  at  a  reduced  spatial
resolution, and  there is an  obvious trade-off between  resolution and
sensitivity.

Although the  CCDs and CMOS  detectors are linear, the  speckle signal
processing  involves  two  nonlinear   operations:  clipping  and  PS
calculation. The PS  is proportional to the square of  the signal, and
an additive background such as CIC events is no longer additive in the
PS.  A  numerical experiment was  conducted to evaluate the  effect of
signal processing on  the relative photometry of a binary  star with a
magnitude difference of 1 mag.  The result is reassuring.  Even in the
conditions   of   SNR$<$1,   when   the  fringes   are   barely   seen
(Figure~\ref{fig:binary}),  no  systematic   trend  in  the  estimated
$\Delta  m$  is   present  for  both  an  EM  CCD   and  a  qCMOS.   In
Figure~\ref{fig:errors},  the  input  $\Delta  m  =  3$  mag  is  also
retrieved without  bias.  A  minor bias  is expected at  $\Delta m
  \approx  0$  because  the   measured  fringe  contrast  $B$  becomes
  independent of the flux ratio: ${\rm d}B/{\rm d}r =0$ at $r=1$.

%---------------------------------------------------------
\section{Toward Automated Observations?}
\label{sec:disc}

Speckle observations are a complex  process. It starts with the preparation
of the  program for each  observing run.  Merging all  active projects
into a  common program has  many advantages and increases  the overall
efficiency,  compared to  the classical  by-program telescope  use.  A
common set of calibrators, for example, ensures consistent calibration
for all projects. Brighter binaries can be measured under poor seeing,
when fainter (and higher-priority) targets would not yield useful data
anyway (they can be pointed on  another night with better seeing). So,
the  observing  program  always  includes more  objects  than  can  be
observed on a  given night, maximizing the use  of allocated telescope
time.

The outcome  of each  observation depends  on the  current conditions,
mostly on the seeing and transparency.  Factors such as telescope shake
are important on windy nights, restricting the pointing direction.  So,
the selection of  targets is managed flexibly in  real time, balancing
between priorities,  target visibility, and current  conditions while
optimizing the  telescope slews.  Quick  online calculation of  the PS
helps the observer to evaluate the HRCam performance and to adjust the
target choice accordingly.  For example,  if a source is resolved into
a new tight binary, observation  of a reference star immediately after
is needed for a correct data reduction.

Considering this  inherent complexity of speckle  observations and the
need to make  real-time decisions based on several  variables, it will
be difficult to  substitute an experienced observer  with an automatic
process. The study done here helps this task in several ways, allowing
to replace subjective  evaluation of the observing  conditions and the
data quality with quantitative metrics  such as $N_{\rm ph}$, $S$, and
SNR.  Simulations will help us  to define reasonable performance goals
and  acceptable SNR  for  each target;  this will  set  the stage  for
automating  observations  in the  future.   The  automation should  be
gradual,  starting from  sequencing  routine  actions and  progressing
toward  robotic  operation  under human  supervision.   Given  current
conditions  and constraints,  the observer  can define  a sequence  of
several targets  for which  the data will  be taken  robotically. This
will  speed  up   the  data  acquisition  and   increase  the  overall
efficiency.   On the  other hand,  the  quality of  the data  acquired
robotically may be worse.

Processing of the  observations acquired on a speckle  night starts by
running  a pipeline  that computes  the  PSs and  the associated  data
products,  as described  in \citet{HRCAM}.   Manual inspection  of the
data and fitting binary parameters  is relatively quick; it takes less
than a half of  the telescope time used to acquire  the data, and about as
much  time  is spent  on  the  subsequent  analysis of  the  results.
Experience helps to distinguish  real binary companions from artifacts
and to  resolve difficult cases  (binaries with small  separations and
large contrast and  triples). In principle, a  suitably trained neural
network can  handle this  task if  the large  data volume  makes their
processing by humans prohibitive.

The  results  of  observations  are incorporated  into  the  observing
program and  used to decide on  the next observations. For  example, a
rapid orbital motion calls for  repeated measurements within a year or
even           sooner.            Thus,          the           process
program$\rightarrow$observations$\rightarrow$program is circular, with
a rapid feedback.  This differs from the classical open-ended approach
where the  data analysis  sometimes is  performed several  years after
their acquisition.

%---------------------------------------------------------
\section{Summary}
\label{sec:sum}

The large  volume of HRCam  speckle data accumulated to  date warrants
the  analysis  of their  quality  performed  here. Parameters  of  the
current EM CCD  detector and the photometric  calibration that relates
source  magnitude to  the  number  of detected  photons  allow a  fair
comparison between the actual level of  the speckle signal $S$ and the
SNR  to  those expected  under  typical  seeing conditions.   For  the
latter,  I adopted  a  simplified model  consisting  of two  turbulent
layers  moving at  different speeds.  The parameters of  the model  are
adjusted to match  typical data.  Simulations help us  to quantify the
effects of  finite exposure  time and  spectral bandwidth,  to predict
the dependence of the  SNR on seeing and stellar magnitude,  and  verify
that the algorithm of PS calculation with clipping is quasi-optimal. I
consider here only  the wide-band $I$ filter used  for observations of
faint stars.

The main results of this study are summarized below. 
\begin{itemize}
\item
The photometric zero points of  HRCam corresponding to a detected flux
of one e-~s$^{-1}$ in  the $I$ and $y$ filters are  25.6 and 24.6 mag,
respectively.  The relation of the instrumental $I$ magnitudes to the Gaia
photometry is established.

\item
The IDL  code for  simulating speckle  images, including  the detector
noise, is developed and made available.

\item
In  the  calculation of  the  PS,  signal  clipping is  necessary  for
maximizing sensitivity with both an EM  CCD and a CMOS. For the latter,
the  optimum clipping  threshold  is 3  times  the RON.
Signal  clipping does  not  spoil the  relative  photometry of  binary
stars.

\item
The speckle PS  decreases uniformly at all  frequencies with increasing
exposure  time. However,  in the  wide $I$  band, the  speckle contrast
declines  with spatial  frequency faster  than in  monochromatic light
(Figure~\ref{fig:poly}).   The   level  of   the  speckle   signal  is
quantified  by $S$,  the decimal  logarithm of  the PS  at half  of the
cutoff frequency.   Under good conditions,  $S \approx -4$ in  the $I$
filter.

\item
Increasing the exposure time from 30 to 60\,ms does not increase
the SNR in the PS but doubles the acquisition time. For faint stars,
it is better to acquire a larger number of short-exposure frames.  

\item
Measurements of faint  binary stars are possible at  SNR $\sim$1. With
the current EM CCD, this corresponds to the $I$ magnitude from 11.5 to
14    mag    for    the    seeing   from    1\farcs4    to    0\farcs6
(Figure~\ref{fig:snr}).   A sensitivity  gain  of at  least  1 mag  is
expected with  a low-noise  CMOS detector  (Figure~\ref{fig:snr2}).  A
CMOS detector overcomes two major limitations  of an EM CCD, namely the
CIC background and the amplification noise.

\item
Quantitative characteristics  of the  HRCam performance  and relations
between seeing conditions,  SNR, and measurement errors  will help to
plan  and execute  speckle observations  and, eventually,  to automate
them in the future.

%\item
%\item

\end{itemize}

Although  this study  is  devoted to  the HRCam  imager  at SOAR,  its
results  may be  useful for  optimizing the performance  of other  speckle
instruments. For example, speckle observations in a wide band on a 4 m
telescope are discussed by \citet{Clark2024a}.  An upgrade from EM CCD
to CMOS is recommended for the current and future speckle imagers.

\begin{acknowledgments} 
The Andor EM CCD camera was loaned to SOAR in 2017 by N.~Law from
the  University   of  North  Carolina,  resulting   in  a  significant
sensitivity gain compared to the previously used detector. I thank
B.~Mason and the anonymous Referee for useful comments on this manuscript.
\end{acknowledgments} 

\appendix

%---------------------------------------------------------
\section{Simulation Code}
\label{sec:sim-alg}

\begin{figure*}[ht]
\epsscale{0.8}
\plotone{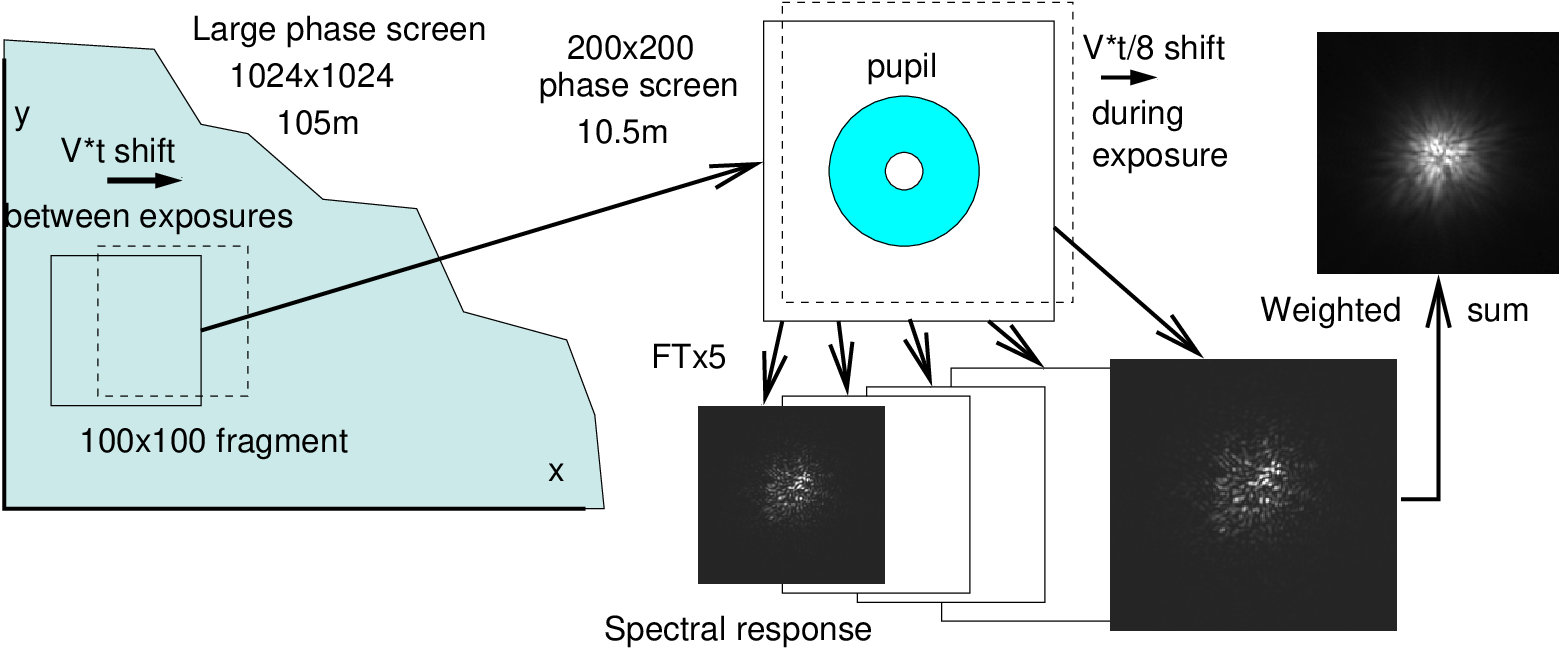}
%\plotone{Code.eps}
\caption{Scheme of the speckle simulation code (see text).
\label{fig:code} 
}
\end{figure*}

The IDL  code {\tt  simspec4.pro} simulates  a realistic  speckle data
cube for the  4.1 m SOAR telescope.  The monochromatic image  of a star at
wavelength  $\lambda$ is  computed  in the standard  way as  a square
  modulus of the FT of the light-wave amplitude at the telescope pupil
  \citep[e.g.][]{Roddier1981,Goodman1985}. The size of the computing
grid   $N$,   200$\times$200,  and   the   angular   pixel  scale   $p
=0\farcs01575$, are chosen to match  the real data.  The physical size
of the grid in the pupil space equals $L = \lambda/p$ (10.5\,m for 0.8
$\mu$m, or 0.05\,m per spatial pixel).   This sets the pupil radius in
pixels;  the   central  obstruction    (0.25   fraction  of  pupil
  diameter)  is  also  emulated   in  constructing  the  pupil  mask.
Figure~\ref{fig:code} helps to visualize the geometry of simulations.

The   random  atmospheric   phase   perturbations obeying
Kolmogorov  statistics  \citep{Roddier1981}  are generated  using  the
standard Fourier  method \citep{Lane1992},  given the  Fried parameter
$r_0  = 0.98  \lambda/\epsilon$  for the  seeing  $\epsilon$   (in
  radians).  In  the   simplest  version  of  the   code,  the  same
pupil space grid was used to  compute the phase screens.  However,
to  simulate the effects  of the  finite  exposure, the  phase screens  of
1024$\times$1024 size  with twice larger pixels  (physical screen size
105\,m) are generated.  The pupil grid corresponds to a 100$\times$100
pixel fragment of the large screen,  and it is interpolated on a finer
200$\times$200 grid for the image calculation.

To  simulate  temporal  evolution  of the  speckle  pattern,  at  each
successive  frame, the  origin of  the selected  fragment of  the large
phase screen is translated horizontally (in  $X$) by $V t$, where $V$ is
the wind speed  and $t$ is the exposure time  (e.g.  1.2\,m for $V=40$
m~s$^{-1}$ and $t=0.03$\,s).  The translation is rounded to an integer
number  of  pixels.   Orthogonal  shifts are  applied  at  every  10th
translation  in  order  to  sample  the  full  large  screen,  so  the
fragment's motion is in fact tilted with respect to the $X$ axis by 1/10
radian (5\fdg7).   When the   selected fragment reaches the  edge of
the  large  screen, it  ``rolls  over''  in both  coordinates  without
discontinuity (the  phase screens are  doubly periodic owing  to their
generation method).

The phase perturbation  at the telescope pupil is  produced by several
turbulent  layers  moving  at  different speeds,  and  the  temporal
evolution of the speckle is governed  mostly by the changing  phase
sum rather  than by the overall  translation over the pupil.   So, two
phase screens are generated.  One is translated in $X$ with a speed of 8
m~s$^{-1}$,  another  in  $Y$  with  a  speed  of  40  m~s$^{-1}$.   The
turbulence  intensity is  equally distributed  between these  screens.
For a seeing  of 0\farcs8, this layout corresponds  to the atmospheric
time  constant   \citep[see definition  in][]{Roddier1981}    of
1.3\,ms  at  500\,nm.   All  parameters  in the  code  can  be  easily
modified.

To reproduce the  image smearing during exposure, the  latter is split
into 8  steps.  The  two phase-screen fragments  selected for  a given
exposure and re-binned on a 200$\times$200 grid are shifted during the
exposure by small  steps of $V t/8$ in the  orthogonal directions.  At
each step, the speckle image is  computed, and each frame of the image
cube is the  average of these eight images.  For  the following frame,
new fragments are cut out from the large phase screens.

Simulation  of speckles  in  a  wide bandwidth  adds  another layer  of
complexity.   The spectral  response of  HRCam  in the  $I$ filter  is
modeled by  a combination  of 5 wavelengths  $\lambda_i$ from  0.75 to
0.90\,$\mu$m  with relative  weights of  [0.724, 0.694,  0.539, 0.347,
  0.190]  --- product  of  the filter  transmission  and the  detector
response    \citep[see  Figure  2  in][]{HRCAM}.   The  effective
wavelength  is  0.822\,$\mu$m.  The  actual  response  depends on  the
spectrum of the star, of course.

For each of  the 8 subframes, we compute five  images corresponding to
$\lambda_i$  and combine  them with  relative weights  defined by  the
spectral  response.  So,  calculation  of one  speckle frame  requires
8$\times$5=40 FTs,  and the simulation  of the image cube  takes about
45\,s.   The  wave  front  distortion   $\Delta  l$  in  linear  units
corresponds to a wavelength-dependent phase  shift of $\Delta \phi = 2
\pi \Delta  l/\lambda$. So,  the phase  perturbation, defined  for the
reference  wavelength  $\lambda_0$,  is  scaled by  the  factor  $a  =
\lambda_0/\lambda_i$.   However, this  is not  sufficient because  the
pupil size and the computing grid are dimensioned for $\lambda_0$, not
for   $\lambda_i$.   This   is  accounted   for  by   stretching  each
monochromatic image  by the factor  $1/a$ (at longer  wavelengths, the
speckles  become  larger).   The  same effect  could  be  achieved  by
shrinking the  phase pattern and  the pupil  function by a  factor $a$
before calculating  the image  by FT.   As a  check, suppose  that the
phase  aberration is  linear (a  pure tilt).   Scaling it  by $a  < 1$
reduces  the tilt  and moves  the image  closer to  the field  center.
Stretching the  image by  $1/a$ times  moves it  back to  the original
position, so  the tilt becomes  achromatic, as expected.   Speckles in
the simulated  polychromatic images are  extended radially, as  in the
real images, while the image moves as a whole owing to random tilts.

The noiseless image cube (either simulated or real) serves as input to
the  general-purpose noise  simulator  {\tt  noisesimul.pro}.  As  the
simulator does  not ``know''  the wavelength, telescope  diameter, and
pixel scale, the  cutoff frequency $f_c$ (in pixels)  must be provided
at input. The  number of photons per frame $N_{\rm ph}$  sets the detected
flux from the  star.  The noise simulation algorithm is  described below; it
is implemented by the following piece of IDL code relevant to one frame:
\begin{verbatim}
  tmp = cube[*,*,i] ; noiseless speckle image
  if keyword_set(binary) then tmp += binary[2]*shift(tmp, binary[0], binary[1])
  tmp = tmp/total(tmp)*Nph ; normalize by average photon number
  for k=0,nx-1 do for l=0,nx-1 do begin ; loops over nx*nx pixels
    npix =  randomn(seed, poisson=tmp[k,l]+CIC)  ; number of events in pixel
    if (ampl gt 1) then begin ; EM CCD, amplification noise
      s = 0 & if (npix gt 0) then for j=0,npix-1 do s += -alog(randomu(seed))
      ncube[k,l,i] = ampl*s + ron*randomn(seed) ; EM CCD
    endif else ncube[k,l,i] = npix + ron*randomn(seed)  ; CMOS
  endfor  ; pixel loop
\end{verbatim}

Each frame  {\tt tmp} is  normalized to a  unit sum and  multiplied by
$N_{\rm ph}$ to  get the expected photon numbers  in each pixel.
This number  serves to  generate the actual  random number  of photons
{\tt npix} after adding the CIC probability (the Poisson random number
generator is  used). For a CMOS  detector, CIC=0, and the  pixel value
equals  {\tt  npix} plus  a  Gaussian  RON, which  must  be
specified in electrons.  For an EM CCD, each photon is ``amplified'' to
an  average level  {\tt  Ampl} (in  ADU)  with a  negative-exponential
distribution, and the resulting pixel signal  is a sum of all randomly
amplified photons.  This   simulates the amplification noise. The
readout  noise is  also  simulated, but  for  an EM  CCD  it should  be
specified in  ADUs rather than in  electrons.  The amplification
step is skipped  for a CMOS by setting {\tt  Ampl=1}.  Simulated noisy
speckle  images  are   illustrated  in  Figure~\ref{fig:frames}.   The
speckle       simulation       code        is       available       at
\url{https://www.ctio.noirlab.edu/~atokovin/speckle/simulation.html} and on Zenodo \citep{specklesimul}.

\facility{SOAR}

%\begin{thebibliography}{99}
%\expandafter\ifx\csname natexlab\endcsname\relax\def\natexlab#1{#1}\fi
%\providecommand{\dodoi}[1]{doi:~\href{http://doi.org/#1}{\url{#1}}}

\bibliography{soar.bib}
\bibliographystyle{aasjournal}

%\clearpage
%\landscape

%\LongTables
%\input{tab6}
%\input{tab7}

\end{document}